\documentclass[10pt,aps,twocolumn,prl,nofootinbib,noshowkeys,superscriptaddress,floatfix]{revtex4-1}

\usepackage{amsmath,amsfonts,amsthm,amssymb}
\usepackage[dvips]{graphics,graphicx}
\usepackage{nicefrac}
\usepackage[usenames,dvipsnames]{color}
\definecolor{darkblue}{RGB}{0,0,196}
\definecolor{darkgreen}{RGB}{0,120,0}
\usepackage[colorlinks=true,linktocpage=true,linkcolor=darkblue,citecolor=red,urlcolor=darkblue]{hyperref}
\usepackage{cancel}
\usepackage{multirow}
\usepackage{longtable}
\usepackage{color}
\usepackage[normalem]{ulem}
\usepackage{hyperref}
\usepackage{bigints}
\usepackage{yfonts}
\usepackage{xparse}
\usepackage{physics}
\usepackage{verbatim}
\usepackage{minibox}
\usepackage{comment}
\usepackage{appendix}

\begin{document}

\preprint{}

\title{Quasiparticle Cosmology}

\author{Amaresh Jaiswal}
\email{a.jaiswal@niser.ac.in}
\affiliation{School of Physical Sciences, National Institute of Science Education and Research, An OCC of Homi Bhabha National Institute, Jatni-752050, India}

\date{\today}


\begin{abstract}

We consider thermodynamics of the Universe within a quasiparticle approach where the collective dynamics of a system is governed by the thermal mass of the constituents. The spacetime dependence of this thermal mass leads to a negative contribution to the pressure and a positive contribution to the energy density, similar to the effect due to dark energy. We propose a mechanism based on thermodynamic arguments to quantify this contribution from the thermal vacuum of the system. For a sufficiently large spacetime variation of the thermal mass, the effective pressure can become negative and could mimic a dark energy equation of state. We validate our framework using results from renormalizable interacting scalar field theory and demonstrate an application on QCD axion. 

\end{abstract}


\maketitle



{\bf \textit{Introduction}}--There exists compelling observational evidence indicating the presence of a small but non-zero value of Einstein's cosmological constant, $\Lambda$, or a component of the material content of the Universe that behaves in a similar manner, exhibiting slow variations across both time and space. The cosmological constant was originally introduced to accommodate static and homogeneous solutions within Einstein's equations in the presence of matter \cite{Einstein:1931}. However, with the discovery of the Universe's expansion \cite{Hubble:1929ig}, its original purpose became unnecessary. Subsequently, there were instances where a nonzero cosmological constant was proposed as an explanation for certain observations, only to be withdrawn when the supporting observational evidence diminished. Until the late 1990s, the prevailing consensus among most physicists was to set the cosmological constant to zero. However, this perspective changed with the discovery in 1998 that the expansion of the Universe is accelerating~\cite{SupernovaCosmologyProject:1998vns, SupernovaCosmologyProject:1997zqe, SupernovaSearchTeam:1998fmf, SupernovaSearchTeam:1998bnz}. This finding suggested the possibility of a positive value for the cosmological constant.

In the meantime, particle theorists recognized that the cosmological constant could be interpreted as a measure of the energy density of the vacuum. This energy density seems to arise from various, apparently unrelated contributions, each with magnitudes significantly larger than the current upper limits on the cosmological constant. Consequently, a fundamental question emerged: why is the observed vacuum energy so exceedingly small compared to the scales of particle physics? This intriguing puzzle regarding the mechanism which suppresses $\Lambda$ extremely precisely to yield an observationally accessible value has gained considerable attention; see Refs.~\cite{Weinberg:1988cp, Felten:1986zz, Carroll:1991mt, Carroll:2000fy, Peebles:2002gy, Padmanabhan:2002ji} for extensive reviews. In this article, we propose a mechanism, based on thermodynamic arguments, to quantify the contribution from thermal vacuum to the energy density corresponding to the cosmological constant. This is achieved by requiring that the collective evolution of a system does not violate fundamental thermodynamic relations. We validate our framework against interacting scalar field theory and apply it to a system of QCD axion with its thermal mass evaluated from effective field theory and lattice QCD calculations. Throughout the text we employ natural units where $ \hbar = c = k_B =1$ and use $(+,-,-,-)$ for the metric signature. 



{\bf \textit{Cosmological constant and vacuum energy}}--The Einstein equation for gravitation in presence of the cosmological constant term is given by
\begin{equation}\label{einstein_eqn}
R_{\mu\nu} - \frac{1}{2}g_{\mu\nu} R - \Lambda\, g_{\mu\nu} = 8\pi G_N  T_{\mu\nu}^{\rm mat},
\end{equation}
where, $g_{\mu\nu}$ is the spacetime metric with corresponding Ricci tensor $R_{\mu\nu}$ and Ricci scalar $R$. Here $G_N$ is the Newton constant, $T_{\mu\nu}^{\rm mat}$ is the energy-momentum tensor of the matter\footnote{The term ``matter" should be interpreted in a broad sense, meaning not only physical matter but also radiation and other relevant quantities.} content of the Universe and $\Lambda$ is the Einstein's cosmological constant. The above equation can be rewritten as
\begin{equation}\label{einstein_eqn_vac}
G_{\mu\nu} = 8\pi G_N \left( T_{\mu\nu}^{\rm mat} +\rho_\Lambda\, g_{\mu\nu} \right),
\end{equation}
where $G_{\mu\nu} \equiv R_{\mu\nu} - \frac{1}{2}g_{\mu\nu} R$ is the Einstein tensor and $\rho_\Lambda \equiv \Lambda/\left( 8\pi G_N \right)$ is attributed to the energy density of the vacuum, as demonstrated below. The l.h.s. of the above equation represents the geometry of spacetime and r.h.s. represents the sum of material content of the Universe and contribution from vacuum energy. It is important to note that the current astrophysical data is consistent with a small non-zero value of the cosmological constant, $\Lambda\simeq 10^{-52}\,\rm{m}^{-2}$. From this, one can estimate 
\begin{equation}\label{cos_eng_den}
\rho_\Lambda\simeq 2.5\times 10^{-11}\,\rm{eV}^4
\end{equation}
to be the value of energy density corresponding to cosmological constant, constrained by the current astrophysical observations.

In order to understand the origin of $\rho_\Lambda$, lets consider the action for a single scalar field $\phi$ with potential $V(\phi)$~\cite{Carroll:2000fy},
\begin{equation}\label{scalar_action}
S = \int d^4 x \sqrt{-g} \left[ \frac{1}{2} g^{\mu\nu}\left(\partial_\mu\phi\right) \left(\partial_\nu\phi\right) - V(\phi) \right],
\end{equation}
where, $g$ is the determinant of the metric tensor. The energy momentum tensor corresponding to the above action can be written as
\begin{eqnarray}\label{scalar_EMT}
T_{\mu\nu} = \frac{1}{2}\!\left(\partial_\mu\phi\right)\!\left(\partial_\nu\phi\right) - \frac{1}{2}\!\left(\partial^\rho\phi\right)\!\left(\partial_\rho\phi\right)g_{\mu\nu} + V(\phi) g_{\mu\nu}.
\end{eqnarray}
The lowest energy configuration in this theory will correspond to vanishing kinetic energy or derivative term, i.e., $\partial_\mu\phi=0$. Therefore, the energy momentum tensor for this configuration reduces to $T_{\mu\nu}^{(0)}=V(\phi_0)g_{\mu\nu}$ where $\phi_0$ minimizes $V(\phi)$. Moreover, the minimum of potential need not be zero and hence $V(\phi_0)$ can in principle be non-vanishing. Consequently, the vacuum energy-momentum tensor can, in general, be written as
\begin{equation}\label{vacuum_EMT}
T_{\mu\nu}^{(0)} = \rho_{\rm vac}\,g_{\mu\nu}.
\end{equation}
The above form of the energy-momentum tensor is equivalent to the term with $\rho_\Lambda$ in Eq.~\eqref{einstein_eqn_vac}, and hence provides the motivation for identification of the cosmological constant with the vacuum energy.

This identification of the cosmological constant with the vacuum energy, however, leads to a very puzzling scenario. The net cosmological constant can be understood as the combination of several seemingly unrelated contributions arising from scalar fields, zero-point fluctuations of each field theory and an inherent bare cosmological constant. Lets consider the vacuua of the standard model field theories. Based on energy scales of various symmetry breaking in the early Universe, one can calculate the vacuum energy densities as
\begin{align}
\rho_{\rm vac}^{\rm QCD} &\sim \left( 0.3\,{\rm GeV} \right)^4 \sim 10^{34}\,{\rm eV}^4, \label{rho_vac_QCD}\\
\rho_{\rm vac}^{\rm EW} &\sim \left( 200\,{\rm GeV} \right)^4 \sim 10^{45}\,{\rm eV}^4,\label{rho_vac_EW}\\
\rho_{\rm vac}^{\rm Pl} &\sim \left( 10^{18}\,{\rm GeV} \right)^4 \sim 10^{108}\,{\rm eV}^4, \label{rho_vac_Pl}
\end{align}
corresponding to QCD, electroweak and Planck scales, respectively. From Eqs.~\eqref{cos_eng_den} and \eqref{rho_vac_Pl}, we see that the energy density corresponding to the observationally constrained cosmological constant, $\rho_\Lambda$, differs by roughly $120$ orders of magnitude from theoretical values of $\rho_{\rm vac}$. It is unrealistic to believe that all large and seemingly unrelated contributions to the vacuum energy cancel precisely to leave a small residual value given in Eq.~\eqref{cos_eng_den}. This is the famous ``cosmological constant problem" or ``vacuum catastrophe"\footnote{Arguments based on anthropic principle have been put forward as a solution to this problem~\cite{Hogan:1999wh}.}. In the following, we propose a mechanism to extract thermal vacuum contribution to $\rho_\Lambda$ based on fundamental thermodynamic arguments.



{\bf \textit{Thermal mass and quasiparticle thermodynamics}}--It is important to note that the thermodynamics and hence the collective dynamics of a system is governed by the thermal mass of the constituents which can depend on medium properties, i.e., temperature and chemical potential of the medium. Interestingly, the concept involving dark matter particles with masses that grow as the Universe expands have been explored earlier~\cite{Casas:1991ky, Garcia-Bellido:1992xlz, Anderson:1997un}; see also Refs.~\cite{Hu:1998kj, Anisimov:2008dz}. Here we consider the statistical mechanics of a system of quasiparticles with temperature dependent mass which, in turn, is spacetime dependent due to evolution of the system. The dispersion relation for a relativistic particle of energy $\omega^*$ and momenta $k$ in such a system can be written as
\begin{equation}\label{disp_rel}
\omega^* (k,T) = \sqrt{k^2 + m^2(T)}.
\end{equation}
The well known expressions for pressure and energy density for an ideal gas can be calculated as~\cite{Landau:1980mil}
\begin{align}
P_{\rm id}(T) &= \pm\frac{gT}{2\pi^2}\int_0^\infty dk\, k^2\,\ln\left[ 1\pm\exp(-\omega^*/T) \right], \label{Pid}\\
\rho_{\rm id}(T) &= \frac{g}{2\pi^2}\int_0^\infty dk\, \frac{k^2\, \omega^*}{\exp(\omega^*/T)\pm 1} , \label{rhoid}
\end{align}
respectively, where, $g$ is the degeneracy factor of the particles and the upper (lower) sign is for fermions (bosons). However, due to temperature dependence of the the particle mass in Eq.~\eqref{disp_rel}, expressions in Eqs.~\eqref{Pid} and \eqref{rhoid} does not satisfy the fundamental thermodynamic relation 
\begin{equation}\label{thermodynamic_rel}
\rho(T) = T\,\frac{dP(T)}{dT} - P(T).
\end{equation}
In earlier studies involving dark matter particles having variable masses, it was indeed discovered that the above thermodynamic relations are not satisfied and a mechanism termed as ``dark entropy production" was proposed in order to resolve this issue~\cite{Casas:1991ky, Garcia-Bellido:1992xlz}. On the other hand, it is expected that the thermodynamic consistency is not guaranteed for a system of particles with variable masses. Therefore, it is important to construct a consistent thermodynamic framework for a system of quasiparticles with spacetime dependent mass.

In order to construct such a framework, we start with an effective Hamiltonian for our statistical system \cite{Gorenstein:1995vm},
\begin{equation}\label{Heff}
H_{\rm eff} = \sum_{i=1}^{g}\sum_{\bf k}\omega^*(k,T)\,a^\dagger_{{\bf k},i}\,a_{{\bf k},i} \,+\, E_0^*,
\end{equation}
where the index $i$ corresponds to particle internal degrees of freedom or degeneracies, and, $a^\dagger_{{\bf k},i}$ and $a_{{\bf k},i}$ are creation and annihilation operators, respectively. Here, $E_0^*$ is the vacuum or zero point energy of the system. In the standard case of temperature independent particle mass, $E_0^*=E_0$ is a constant which is divergent and is usually removed from the system's energy spectrum. However, for dispersion relation of the form given in Eq.~\eqref{disp_rel}, $E_0^*$ becomes a function of temperature and can not be eliminated. Using the effective Hamiltonian in Eq.~\eqref{Heff}, one can calculate the thermodynamic pressure and energy density by recognizing the quasiparticle number operator $N=\sum_{i=1}^{g}\sum_{\bf k}\,a^\dagger_{{\bf k},i}\,a_{{\bf k},i}$ and using the definitions~\cite{Landau:1980mil}
\begin{align}
P(T) &= \frac{T}{V}\,\ln {\rm Tr} \left( e^{-H_{\rm eff}/T} \right) \equiv \frac{T}{V}\,\ln Z_{\rm eff}(T,V),  \label{press_def}\\
\rho(T) &= \frac{1}{V}\, \frac{1}{Z_{\rm eff}(T,V)}\, {\rm Tr} \left( H_{\rm eff}\,e^{-H_{\rm eff}/T} \right). \label{eng_den_def}
\end{align}
In the thermodynamic limit, i.e., for volume $V\to\infty$, the summations in Eq.~\eqref{Heff} reduces to integration over the momentum phase-space,
\begin{equation}\label{sum_int}
\sum_{i=1}^{g}\sum_{\bf k} ~\to~ V\frac{g}{(2\pi)^3}\int d{\bf k}.
\end{equation}
Using the above identification in Eqs.~\eqref{press_def} and \eqref{eng_den_def}, we obtain
\begin{equation}\label{P_rho_B*}
P(T) = P_{\rm id}(T) -B^*(T), \quad
\rho(T) = \rho_{\rm id}(T) + B^*(T),
\end{equation}
where, $\displaystyle{B^*\equiv\lim_{V\to\infty}E_0^*/V}$. 

The function $B^*(T)$ in the above equation is defined by imposing the thermodynamic self-consistency. Note, however, that the above expressions for $P(T)$ and $\rho(T)$ are divergent because the term $B^*(T)$ also contains the diverging contribution from the constant zero point energy $E_0$. In order to eliminate the divergent part, we define $B(T)=B^*(T)-B^*(T=0)$, which is finite. We shall demonstrate shortly that this renormalized $B(T)$ contributes to restoring the thermodynamic consistency of the system. As done in the standard case with constant mass, we remove the divergent contribution $B^*(T=0)$ from Eq.~\eqref{P_rho_B*} to get
\begin{align}
P(T) &= P_{\rm id}(T) -B(T), \label{P_fin}\\
\rho(T) &= \rho_{\rm id}(T) + B(T). \label{rho_fin}
\end{align}
Demanding that the above expressions for $P(T)$ and $\rho(T)$ satisfies the thermodynamic relation, Eq.~\eqref{thermodynamic_rel}, one arrives at the condition~\cite{Gorenstein:1995vm, Jeon:1994if, Romatschke:2011qp, Tinti:2016bav, Czajka:2017wdo}
\begin{eqnarray}\label{dBdT}
\frac{dB}{dT} = -\frac{gm}{2\pi^2}\,\frac{dm}{dT}\!\!\int_0^\infty \!\!\!\! \frac{k^2 dk}{\omega^*(k,T)}\, \frac{1}{\exp\!\left[ \omega^*(k,T)/T \right]\pm 1},
\end{eqnarray}
where, we recall again that the upper (lower) sign is for fermions (bosons). 

We perform the momentum integral in Eq.~\eqref{dBdT} to get
\begin{equation}\label{dBdT_fin}
\frac{dB}{dT} = -\frac{g\,m^2\,T}{2\pi^2}\,\frac{dm}{dT} \sum_{l=1}^\infty (\mp 1)^{l-1} \frac{K_1\left(l\,m/T\right)}{l} ,
\end{equation}
where $K_1$ is the modified Bessel function of second kind. Integrating up from absolute zero to the temperature $T_0$, and keeping in mind that $B(T=0)=0$, we get
\begin{equation}\label{B(T)}
B(T_0) = -\frac{g}{2\pi^2}\!\sum_{l=1}^\infty (\mp 1)^{l-1} \!\! \int_0^{T_0} \! \left[ \frac{m^2 T}{l}\, \frac{dm}{dT}\, K_1\!\left(l\,m/T\right) \right] \! dT .
\end{equation}
It is important to note that the condition $B(T=0)=0$ corresponds to \emph{zero temperature renormalization} of the thermal vacuum.

To see the effect of the modifications of pressure and energy density on the evolution of the Universe, we note that the energy-momentum tensor of a non-dissipative fluid consisting of particles with constant mass can be written as
\begin{equation}\label{Tmunu_id}
T^{\mu\nu}_{\rm id} = \left(\rho_{\rm id}+P_{\rm id}\right) u^\mu u^\nu - P_{\rm id}\, g^{\mu\nu},
\end{equation}
where $u^\mu$ is the fluid four-velocity. According to Eqs.~\eqref{P_fin} and \eqref{rho_fin}, the energy momentum tensor of the quasiparticle system takes the form
\begin{equation}\label{Tmunu_full}
T^{\mu\nu} = T^{\mu\nu}_{\rm id} + B(T)\,g^{\mu\nu}.
\end{equation}
Comparing the r.h.s. of the above equation with that of Eq.~\eqref{einstein_eqn_vac}, we find that $B(T)$ is equivalent to $\rho_\Lambda$. It is important to note that this framework provides a mechanism to quantify thermal vacuum contribution to $\rho_\Lambda$ required to establish thermodynamic consistency of a system of quasiparticles. In the following, we verify our framework in the context of interacting scalar field theory.



{\bf \textit{Interacting scalar field theory}}--Let us consider an interacting scalar field theory action, given in Eq.~\eqref{scalar_action}, with the potential given by $\displaystyle{ V(\phi) = \frac{1}{2} m^2 \phi^2 + \frac{1}{4} \lambda \phi^4 }$. In the weak coupling limit, the pressure can be obtained from free energy $P(T)=-f(T)$, in a simplified form for vanishing bare mass. One can also calculate the effective thermal mass, $m_{\rm eff}(T)$, in this case. Up to linear order in the coupling constant, one obtains~\cite{Laine:2016hma},
\begin{equation}\label{pphi_meff}
P(T) = \frac{\pi^2\, T^4}{90} - \frac{\lambda_R\, T^4}{192}, \quad m^2_{\rm eff}(T)=\frac{\lambda_R\, T^2}{4},
\end{equation}
where $\lambda_R$ is the renormalized coupling constant. Using the expression of $m_{\rm eff}(T)$, we verify in the following that the present framework, based on quasiparticle picture, leads to the expression of $P(T)$ given above.

We first evaluate $P_{\rm id}$ using Eq.~\eqref{Pid} for the case of bosons. We employ $m_{\rm eff}(T)$ of Eq.~\eqref{pphi_meff} in the energy-momentum dispersion relation, Eq.~\eqref{disp_rel}, to evaluate $P_{\rm id}$ up to linear order in $\lambda_R$. We also evaluate $B(T)$ up to linear order in $\lambda_R$ using $m_{\rm eff}(T)$ from Eq.~\eqref{pphi_meff} in Eq.~\eqref{B(T)}. After performing the integrals, we obtain
\begin{equation}\label{pid_B}
P_{\rm id} (T) = \frac{\pi^2\,T^4}{90} - \frac{\lambda_R\,T^4}{96}, \quad B(T) = - \frac{\lambda_R\,T^4}{192}.
\end{equation}
Using the above expressions in Eq.~\eqref{P_fin}, we find that the expression for $P(T)$ matches exactly with that given in Eq.~\eqref{pphi_meff}. Hence we conclude that the thermodynamics of a renormalizable quantum field theory can be established if the effective thermal mass is known. While we have verified the validity of the quasiparticle framework in the weak coupling limit, we conjecture that it holds more generally for an interacting quantum field theory. Based on this validation of the quasiparticle approach from a more consistent quantum field theoretic picture, we propose that $B(T)$ quantifies the thermal vacuum contribution to $\rho_\Lambda$. This may prove to be useful in addressing the fine tuning problem of the cosmological constant.



{\bf \textit{Thermodynamics of QCD axion}}--As an application of the proposed framework, we consider a system of QCD axion. The axion was originally introduced as a dynamic solution to the strong CP (charge conjugation and parity) problem of QCD~\cite{Peccei:1977hh, Peccei:1977ur, Wilczek:1977pj, Weinberg:1977ma}. The QCD axion continues to be a highly promising candidate for dark matter~\cite{Abbott:1982af, Preskill:1982cy, Dine:1982ah, Turner:1989vc, Addazi:2022whi} and therefore its mass is of central importance~\cite{Vysotsky:1978dc}. Current efforts are underway to search for axion dark matter~\cite{Rosenberg:2015kxa}, and there are proposals for future experiments~\cite{IAXO:2019mpb, Bahre:2013ywa} that hold the potential to significantly expand the search window~\cite{Graham:2015ouw}. In order to study the axion thermodynamics, we require the axion thermal mass which is determined by the QCD topological susceptibility~\cite{Petreczky:2016vrs, Dine:2017swf}.

\begin{figure}[t!]
    \centering
    \includegraphics[width=\linewidth]{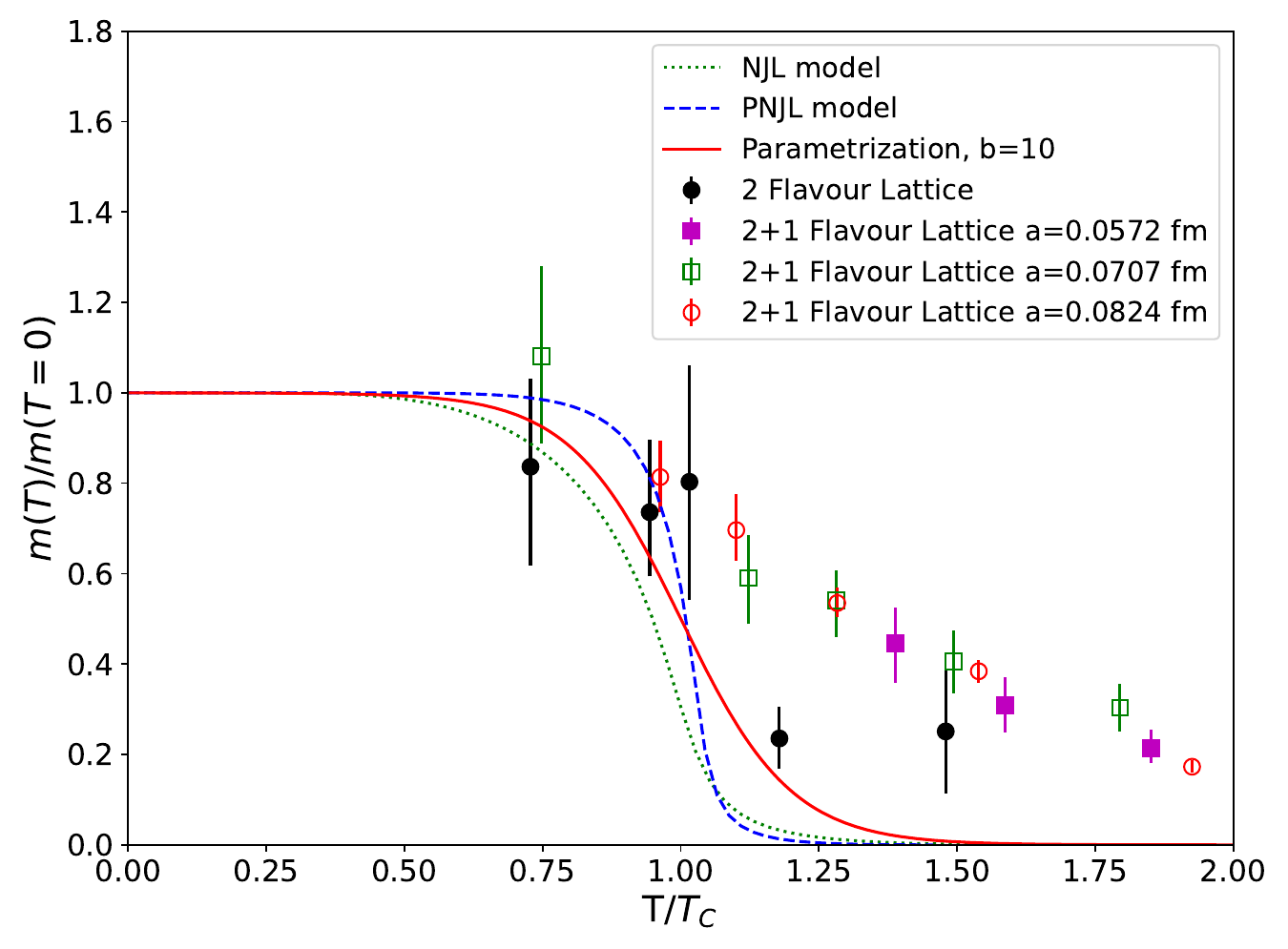}
    \vspace{-6mm}
    \caption{Variation of axion mass with temperature for NJL, PNJL and lattice QCD calculations~\cite{Das:2023ltx}. Also plotted is the parametrization used here for the temperature dependence of the axion mass, given in Eq.~\eqref{m(T)}. }
    \vspace*{-2mm}
    \label{fig_1}
\end{figure}

For the temperature dependence of the axion mass, we consider results from Nambu Jona-Lasinio (NJL) model, Polyakov loop extended Nambu Jona-Lasinio (PNJL) model~\cite{Das:2020pjg} and Lattice QCD~\cite{Alles:2006ua}, as shown in Fig.~\ref{fig_1}. To capture the effective behavior of axion thermal mass from these calculations, we employ a parametrization
\begin{equation}\label{m(T)}
m(T) = \frac{m_0}{\exp\left[b\,(T/T_c - 1)\right] + 1},
\end{equation}
where $m_0$ is the axion mass at zero temperature and we take $T_c\sim 188$~MeV. In Fig.~\ref{fig_1}, we also show the parametrization for one representative value of $b=10$. At zero temperature, PNJL model predicts $m_0f_a\simeq 6\times 10^{-3}~{\rm GeV}^2$~\cite{GrillidiCortona:2015jxo}, where $f_a$ is the axion decay constant. Certain approximate bounds can be put on the decay constant from axion-nucleon interaction and cosmological considerations, $10^9\,{\rm GeV} \lesssim f_a \lesssim 10^{11}\,{\rm GeV}$~\cite{Dror:2023fyd}. From here, we obtain $6\times 10^{-3}\,{\rm eV} \lesssim m_0 \lesssim 6\times 10^{-5}\,{\rm eV}$. We use Eq.~\eqref{m(T)} in Eq.~\eqref{B(T)} for the case of bosons, and evaluate $B(T)$ for axions numerically.

In presence of conjectured axion-photon couplings~\cite{Peccei:2006as}, we assume axion to be in thermal equilibrium with Cosmic Microwave Background (CMB) photons. This makes $B(T_0)$ spatially homogeneous due to homogeneity of CMB radiation. Moreover, from Fig.~\ref{fig_1}, we note that at low temperature, i.e. at current epoch, the axion mass is almost constant as a function of temperature with small constant slope. This renders $B(T_0)$ a constant as a function of time as well, at current epoch. These features are consistent with the properties of cosmological constant.

The present CMB temperature of the Universe, $2.73~{\rm K}$, can be expressed in eV units as $T_0=6.6\times10^{-4}$~eV. We consider the axion degeneracy factor as $g=1$. We find that for $m_0=6\times 10^{-4}$~eV, we obtain $B(T_0)\simeq 10^{-30}~{\rm eV}^4$. This value, of course, is quite far from that given in Eq.~\eqref{cos_eng_den} obtained from astrophysical data. However, one can further improve this to $B(T_0)\simeq 10^{-15}~{\rm eV}^4$ by considering a system of quasiparticles whose thermal mass is parametrized as in Eq.~\eqref{m(T)} but with $T_c$ replaced by $T_0$, implying a sharp decrease in the thermal mass. The order of magnitude of values obtained for $B(T_0)$ remains unchanged for $b>10$. Nevertheless, it is important to note that we obtain a small positive value of $B(T_0)$, which is in the right direction towards obtaining the value of $\rho_\Lambda$ consistent with observations. 

We would like to emphasize that the estimation of $B(T_0)$ presented here is just a qualitative demonstration of the application of the proposed framework. The purpose of the above analysis is not to argue that the thermodynamics of QCD axion can provide a phenomenologically viable component of the dark sector. Moreover, the plausibility of this situation would only be evident if this particular component $B(T_0)$ effectively assumes the role of $\rho_\Lambda$ in the standard concordance model. Instead, our purpose in outlining this scenario is solely to provide an example of how the thermodynamical consistency requirement could lead to emergence of a small positive value for $B(T_0)$, which can potentially influence the large-scale dynamics of the Universe.



{\bf \textit{Summary and Outlook}}--To summarize, we have presented a mechanism, based on thermodynamic arguments, to quantify the contribution from the zero-point energy of the system to the evolution of the Universe. We considered evolution of Universe within a quasiparticle approach where the collective dynamics of a system is governed by the spacetime dependent thermal mass of the constituents. The requirement of thermodynamic consistency naturally leads to a contribution from the vacuum energy to $\rho_\Lambda$. We validated our proposed framework against interacting scalar field theory and applied it to study a thermodynamic system of QCD axion.

Looking forward, it will be interesting to apply the proposed framework to study the expansion of the Universe through Friedmann equations. Note that $B(T)$ is temperature dependent which could play a crucial role for expansion in the early Universe when there was rapid temperature evolution. It is also important to consider dissipative effects in the evolution~\cite{Padmanabhan:1987dg, Fabris:2005ts, Gagnon:2011id, Floerchinger:2014jsa}. Further, it will be interesting to compare the predictions of this framework with observational data. We leave these studies, within the quasiparticle model of cosmology, for future work.


\begin{acknowledgements}
Useful discussions with Sayantani Bhattacharyya, Tuhin Ghosh, Sunil Jaiswal, Najmul Haque and Hiranmaya Mishra are gratefully acknowledged. The author thanks Arpan Das and Nachiketa Sarkar for help with data and plot of Fig.~\ref{fig_1}. The author was supported in part by the DST-INSPIRE faculty award under Grant No. DST/INSPIRE/04/2017/000038. 
\end{acknowledgements}

\vspace{-0.1cm}
\bibliography{ref}
\vspace{-0.1cm}

\end{document}